\documentstyle[12pt,graphicx]{article}
\def\beqar {\begin{eqnarray}}
\def\eeqar {\end{eqnarray}}
\def\beq {\begin{equation}}
\def\eeq {\end{equation}}
\def\A{{\cal A}}

\def\C{{\cal C}}

\def\G{{\cal G}}

\def\S{{\cal S}}

\def\D{{\cal D}}

\def\M{{\cal M}}

\def\al{\alpha}
\def\bt{\beta}
\def\del{\delta}

\def\ga{\gamma}

\def\ep{\epsilon}
\def\la{\lambda}
\def\La{\Lambda}
\def\om{\omega}
\def\Om{\Omega}
\def\th{\theta}

\def\si{\sigma}
\def\Si{\Sigma}

\def\d{\partial}

\def\ba{{\bar a}}
\def\bz{{\bar z}}

\def\bom{{\bar \omega}}

\def\hf{\frac{1}{2}}

\def\<{\langle}
\def\>{\rangle}
\def\re{{\rm Re}}
\def\im{{\rm Im}}
\def\Tr{{\rm Tr}}

\def\diag{{\rm diag}}

\begin{document}

\begin{titlepage}
\null\vspace{-62pt} \pagestyle{empty}
\begin{center}
%\rightline{} \rightline{CCNY-HEP-/05}
\vspace{1.0truein}

{\Large\bf On the deconfining limit in \\
\vspace{.3cm}
(2+1)-dimensional Yang-Mills theory} \\

%%%%%%%%%%%%%%%%%%%%%%%%%%%%%%%%%%%%%%%%%%%%%%%%%%%%%%%%%%
\vspace{1.0in} YASUHIRO ABE \\
\vskip .12in {\it Cereja Technology Co., Ltd.\\
3-1 Tsutaya-Bldg.5F Shimomiyabi-cho  \\
Shinjuku-ku, Tokyo 162-0822, Japan}\\
\vskip .07in {\tt abe@cereja.co.jp}\\
\vspace{1.3in}
%%%%%%%%%%%%%%%%%%%%%%%%%%%%%%%%%%%%%%%%%%%%%%%%%%%%%%%%%%%%
\centerline{\large\bf Abstract}
\end{center}
We consider (2+1)-dimensional
Yang-Mills theory on $S^1 \times S^1 \times {\bf R}$
in the framework of a Hamiltonian approach developed by Karabali, Kim and Nair.
The deconfining limit in the theory can be discussed in terms of
one of the $S^1$ radii of the torus ($S^1 \times S^1$),
while the other radius goes to infinity.
We find that the limit agrees with the previously known result for
a dynamical propagator mass of a gluon. We also make comparisons with
numerical data.

\end{titlepage}
%%%%%%%%%%%%%%%%%%%%%%%%%%%%%%%%%%%%%%%%%%%%%%%%%%%%%%%%%%%%%
\pagestyle{plain} \setcounter{page}{2} %\baselineskip =14pt

%%%%%%%%%%%%%%%%%%%%%%%%%%%%%%%%%%%%%%%%%%%%%%%%%%%%%%%%%%%%%%%%%%
\section{Introduction}

The Hamiltonian approach to (2+1)-dimensional Yang-Mills theory has
been known over a decade as a novel framework for non-perturbative
analysis of the theory \cite{KN1}.
Technical elaborations might be necessary in the formulation of the
Hamiltonian approach, in particular, with regard to regularization
processes \cite{KKN1}, but the upshot of the Hamiltonian formulation
is quite simple; namely, it gives rise to (a) an interpretation of an origin
of the mass gap and (b) an analytic calculation of the string tension for
$SU(N)$ gauge groups \cite{KKN1,KKN2}.
These results, obtained by Karabali, Kim and Nair (KKN), are remarkable
not only in consequence of a (conformal) field
theoretic framework but also in comparison with lattice simulations
of the string tension \cite{Teper1}.
Recent developments relevant to the Hamiltonian approach
can be found in \cite{LMY1,Brits,AKN,KN2,Fukuma1}. There are
also various other analytic approaches to the theory; for recent progress,
see, for example, \cite{Orland1,PSR1,Feuchter:2007mq}.

In the present paper, we follow the Hamiltonian approach to consider
the deconfining limit in (2+1)-dimensional Yang-Mills theory.
In the Hamiltonian approach, confining properties of the theory can be
shown by the following steps \cite{KKN2}:
\begin{enumerate}
  \item matrix parametrization of gauge fields;
  \item calculation of a gauge-invariant measure on the configuration space;
  \item evaluation of a vacuum-state wave function $\Psi_0$ and its inner product
  in terms of gauge-invariant variables;
  \item calculation of the vacuum expectation value of the Wilson loop operator
  $\< \Psi_0 | W(C) | \Psi_0 \>$; and
  \item reading off the area law or the positive string tension from
  $\< \Psi_0 | W(C) | \Psi_0 \>$.
\end{enumerate}
The goal of this paper is to rephrase each of the above steps for the theory
on $S^1 \times S^1 \times {\bf R}$ such that we
can discuss the limit of vanishing string tension in terms of
one of the $S^1$ radii which may correspond to a parameter of finite temperature.
For such a parameter we will use $\im \tau$, where $\tau$ is the modular parameter
of a torus $(S^1 \times S^1)$ in the three-dimensional space
$S^1 \times S^1 \times {\bf R}$ of our interest.
A radius corresponding to the other $S^1$ is taken to be large (or
equivalently the corresponding winding number is taken to be large,
with the radius being finite) so that we can identify the theory on
$S^1 \times S^1 \times {\bf R}$
as a planar theory at a finite temperature in such a limit.
This implies that one of the $S^1$ directions corresponds to
the time coordinate.

In the KKN Hamiltonian approach, one takes the temporal gauge $A_0 = 0$
and analyses are made entirely on complexified spatial dimensions by use
of conformal properties.
In the present paper, we however take a different gauge, say $A_x = 0$,
to make an analysis of gauge potentials on torus which include a time component.
Such an analysis may contain
subtlety in discussion of physical dynamics
but for argument of static properties it may still be useful
since the theory of interest is relevant to the one with imaginary
time or the Euclidean metric.
In this paper, leaving that subtlety aside,
we focus rather on construction of the vacuum wave function $\Psi_0$
in pure Yang-Mills theory on $S^1 \times S^1 \times {\bf R}$ (with one
of $S^1$ directions corresponding to an imaginary time coordinate),
following the framework of the KKN Hamiltonian approach.
We shall not deal with a more involved regularization program,
either. Note that (2+1)-dimensional Yang-Mills theory on torus in
spatial dimensions has been studied before
in a different context \cite{OrlSem}.

Apart from what have been mentioned, our main motivation to consider
the deconfining limit is currently available lattice data on
the deconfinement phase transition in (2+1)-dimensional Yang-Mills
theory \cite{Teper2,Teper3,Narayanan1,Narayanan2,Holland1,Holland2}.
In the Hamiltonian approach, the string tension is obtained for a
continuum strong coupling region where $e^2/p \gg 1$ is realized, with
$e$ and $p$ being the coupling constant and a typical momentum scale,
respectively. So we shall limit our analysis in this region and
do not discuss the nature of deconfinement phase transition.
What we aim at is, however, to obtain a critical temperature of
deconfinement transition which can be compared with the numerical data.

Study of (2+1)-dimensional Yang-Mills theory on $S^1 \times S^1 \times {\bf R}$
is therefore physically well-motivated.
In order to execute the study in the Hamiltonian approach, however,
there is a key mathematical concept to be reminded of, that is,
physical states of the planar Yang-Mills theory in the
Hamiltonian approach can be described in terms of
holomorphic wave functionals of Chern-Simons theory.
In the next section, we briefly review this relation.
Once we understand how this relation arises,
consideration of the toric theory would be clearer by use of the so-called
Narashimhan-Seshadri theorem \cite{NaraSesh}, which we also
mention in the next section.

In section 3, following the references \cite{GaweKupi1,BosNair1,BosNair2},
we present a matrix-parametrization of gauge
potentials such that it is incorporated with the zero modes of torus.
This section covers the first two steps of the above list.
In section 4, we deal with the third step.
Technically speaking, the objective of this section is to find
holomorphic functionals of Chern-Simons theory
on torus in a context of geometric/holomorphic quantization
\cite{BosNair1,BosNair2}.
In section 5, utilizing the results of section 4, we
consider the last two steps of the above list. We obtain
an expression for a deconfinement temperature and
make comparisons with lattice data.
In the last section, we present brief concluding remarks.

%%%%%%%%%%%%%%%%%%%%%%%%%%%%%%%%%%%%%%%%%%%%%%%%%%%%%%%%%%%%%%%%%%
\section{Review of the KKN Hamiltonian approach}

In the Hamiltonian approach, the gauge potentials $A_i$
$(i=1,2,3)$ are parametrized
by the elements of $SL(N, {\bf C})$.
The gauge group we consider is $G=SU(N)$;
$A_i$ can be written as $A_i = -i t^a A^{a}_{i}$
where $t^a$'s are the elements
of $SU(N)$ represented by $(N \times N)$-matrices,
satisfying $\Tr(t^a t^b) = \hf \del^{ab}$ and $[t^a, t^b]=i f^{abc}t^c$.
Note that under the temporal gauge $A_0 = 0$ the gauge potentials are
described by $A_z = \hf(A_1 + i A_2)$, $A_\bz = \hf(A_1 -i A_2)$ where
$z=x_1 - i x_2$, $\bz=x_1 + i x_2$ are a complex combination of
the spatial coordinates $(x_1, x_2)$.
Matrix parametrization of the gauge potentials is given by
\beqar
A_z &=&  - \d_z M ~ M^{-1} \nonumber \\
A_\bz &=&  M^{\dag -1}  \d_\bz M^\dag
\label{1}
\eeqar
where $M(z,\bz), M^\dag(z, \bz)$ are the elements of  $SL(N, {\bf C})$.
Gauge transformations of $A_z , A_\bz$ can be realized by
$M \rightarrow g M$, $M^\dag \rightarrow M^\dag g^{-1}$ with $g \in SU(N)$.
A gauge invariant matrix variable is given by $H=M^\dag M$. The parametrization
(\ref{1}) corresponds to step 1 of the list in the introduction.

Let $\A$ denote the set of all gauge potentials $A_{i}^a$.
The gauge-invariant configuration space is then given by
\beq
\C= \A / \G_*
\label{2}
\eeq where
\beq
\G_* = \{ \mbox{set of all $g(\vec{x}):{\bf R}^2 \rightarrow SU(N)$,
 with $g \rightarrow 1$ as $|\vec{x}| \rightarrow \infty$}  \} \,.
\label{3}
\eeq
The calculation of the gauge-invariant measure in step 2 leads to
the following result, up to an irrelevant constant factor \cite{KN1}:
\beq
d \mu(\C) ~=~ d\mu(H) e^{2c_A \S_{WZW}(H)}
\label{4}
\eeq
where $c_A$ denotes the quadratic Casimir of $G$ for the adjoint
representation, $c_A \del^{ab} = f^{amn}f^{bmn}$. For $G=SU(N)$, this
is equal to $N$. $\S_{WZW}(H)$ is
the action for a $G^{\bf C} / G$ Wess-Zumino-Witten (WZW) model
where $G^{\bf C}=SL(N, {\bf C})$ is the complexification of $G=SU(N)$.
Explicitly, the action is given by
\beq
\S_{WZW}(H) = \frac{1}{2 \pi} \int d^2 z \Tr(\d_z H \d_\bz H^{-1})
+ \frac{i}{12 \pi} \int d^3 x \ep^{\mu\nu\al} \Tr( H^{-1}\d_\mu H H^{-1}\d_\nu H
H^{-1}\d_\al H)
\label{5}
\eeq
where $dz^2$ (or $dx^2)$ is a real two-dimensional volume element which
is equivalent to $dzd\bz / 2i$.
For description of a coset model in connection with gauged WZW models,
one may refer to \cite{GaweKupi1,gWZW1,gWZW2,Witten1}.
The inner product of physical states in general, not only for vacuum
states, can be evaluated as correlators of the
$G^{\bf C} / G$-WZW model;
\beq
\<1 | 2\> = \int d\mu(H) e^{2c_A S_{WZW}}\Psi_{1}^* (H) \Psi_{2}(H) \,.
\label{6}
\eeq
where $\Psi^*(H)$ indicates the complex conjugate of a wave functional $\Psi(H)$.
The wave functionals for the {\it vacuum state} can be given by
\beq \Psi_0 (H) =1 \, .
\label{7}
\eeq
Equation (\ref{6}) provides an essential setup for the KKN Hamiltonian
approach.

It is known that current correlators of a WZW model can be
generated by holomorphic wave functionals of Chern-Simons theory
\cite{BosNair1,BosNair2}.\footnote{It is well known by the study of
knots in terms of Chern-Simons theory \cite{Witten:1988hf} that
the conformal blocks of current algebra on $\Si$
(or the current correlators of a WZW model on $\Si$)
correspond to
the sections of holomorphic line bundle over $\M$
(or the generating functional of Chern-Simons theory on $\M_3$), where
$\Si$ is a two-dimensional compact space,
$\M$ is a moduli space of flat connections on a $G$-bundle over $\Si$ and
$\M_3$ is a three-dimensional closed manifold with its boundary being
$\d \M_3 = \Si$.}
The wave functional $\Psi(H)$
can then be interpreted as a functional which arises
from a holomorphic wave functional $\psi[A_\bz]$
of Chern-Simons theory. In a context of geometric/holomorphic quantization
of Chern-Simons theory \cite{BosNair1,BosNair2,ADW},
a polarization condition is imposed on the functional, {\it i.e.},
\beq \Psi[A_\bz] = e^{-\frac{K}{2}} \psi[A_\bz] \,. \label{8} \eeq
Here $K$ is a K\"{a}hler potential associated with
the phase space of Chern-Simons theory in the $A_0 = 0$ gauge;
\beq
K ~=~ \frac{k}{2\pi} \int_\Si A_{\bz}^{a} A_{z}^{a}
~=~ -\frac{k}{\pi} \int_\Si \Tr(A_\bz A_z )
\label{9}
\eeq
where $\Si$ indicates Riemann surface and
$k$ is the level number of Chern-Simons theory.
The integral over $\Si$ is taken for $dz^2 = dz d\bz /2i$.
In terms of the parametrization (\ref{1}),
$\psi[A_\bz]$ can be written as $\psi[M^\dag]$.
The flatness of the gauge potentials, which is required
as the equation motion for $A_0$ (or the gauss law constraint) of
Chern-Simons theory, must be satisfied on the holomorphic
wave functional $\psi[A_\bz]$, {\it i.e.}, $F_{z\bz} \psi[A_\bz]=0$ where
$F_{z\bz}= \d_z A_\bz - \d_\bz A_z + [A_z, A_\bz]$.
This leads to
$\psi[M^\dag] = e^{k \S_{WZW}(M^\dag)}$.
The inner product of the gauge-invariant physical states
is then given by \cite{BosNair2}
\beqar
\<1|2\>_{CS} &=& \int d\mu(\C) e^{-K} \psi_{1}^* \psi_2  \nonumber \\
&=& \int d\mu(H) e^{(2 c_A + k)  \S_{WZW}(H)}
\label{10}
\eeqar
where we use (\ref{4}) and $\psi_{1}^* = \psi_1 [M] = e^{k \S_{WZW}(M)}$
together with
the Polyakov-Wiegmann identity \cite{PW}
\beqar
\nonumber \S_{WZW}(H)=\S_{WZW}(M^\dag M)
&=&\S_{WZW}(M^\dag)+\S_{WZW}(M) -\frac{1}{\pi}\int_\Si
\Tr(M^{\dag -1}\d_\bz M^\dag \d_z M M^{-1}) \\
&=&\S_{WZW}(M^\dag)+\S_{WZW}(M) + \frac{1}{\pi}\int_\Si
\Tr(A_\bz A_z) \,. \label{11}
\eeqar
Comparing (\ref{6}) and (\ref{10}), we find that the inner product of the vacuum
wave functionals on the planer Yang-Mills theory
can be obtained by the inner product (\ref{10}) in the limit of $k \rightarrow 0$.
The theory defined by the correlator (\ref{10}) with positive $k$ is known as
Yang-Mills-Chern-Simons theory \cite{YMCS,Asorey,Cornwall}.

It is interesting that the physical states of (2+1)-dimensional Yang-Mills theory
can be obtained in terms of the holomorphic wave functionals
of Chern-Simons theory. A simple explanation of this relation is that,
as shown in the first reference of \cite{KN1},
under the $A_0 = 0$ gauge commutation rules
among the gauge potentials $A_i$'s and the electric fields $E_i$'s (which are
canonical momenta of $A_i$'s) can be interpreted as
two copies of the Chern-Simons commutation rules among $A_z$'s and $A_\bz$'s
in the same gauge.

For physical states other than the vacuum, the holomorphic wave
functional $\psi[A_\bz]$ of Chern-Simons theory may be expressed
as $\psi[M^\dag] = e^{k S_{WZW}(M^\dag)} F[M^\dag]$ in general where $F[M^\dag]$
is a matrix function of $M^\dag$. Thus $\psi[M^\dag]$ may not lead to the
gauge invariant functional $\Psi(H)$ in (\ref{6}). However, as shown in \cite{KN2},
one can in fact take a suitable gauge choice such that
$F[M^\dag]$ depends on the current of the hermitian $SL(N, {\bf C})/SU(N)$-WZW model,
$J^a = \frac{c_A}{\pi} (\d_z H ~ H^{-1})^a$.
In terms of this gauge-invariant current, the flatness of the gauge potential
corresponds to an equation of motion of the hermitian WZW model, $\d_\bz J^a = 0$.
In the Hamiltonian approach, this indicates the vanishing of magnetic fields which
act on $\psi[A_\bz]$. Note that magnetic fields do not necessarily
vanish when acted on $\Psi[A_\bz]$.
In the present paper, effects of
magnetic fields will not be discussed in constructing the vacuum-state
wave functional.

Now let us return to the toric theory.
It is known that there exist flat connections (or nontrivial gauge potentials that
lead to vanishing curvature) for any compact two-dimensional spaces
$\Si$ which have complex structure.\footnote{In this case,
the moduli space of flat connections on a $G$-bundle over $\Si$
can be identified with
the moduli space of a stable holomorphic $G^{\bf C}$-bundles on $\Si$.
This mathematical fact is known as Narashimhan-Seshadri theorem \cite{NaraSesh}.}
The relation between the current correlators of a WZW model
and the holomorphic wave functions of Chern-Simons theory, given by
the expression of (\ref{6}) or (\ref{10}), therefore holds for any
Riemann surfaces $\Si$ including a torus.
For the study of WZW models on Riemann surfaces and on torus
in particular, see \cite{GaweKupi1,Bernard,FalcGawe1}.
A main purpose of the present paper from a mathematical perspective is to clarify
the structure of the inner product of such holomorphic wave functionals
when the Riemann surface is given by a torus.

As seen in the next section, one can in fact
incorporate zero modes of torus into
a matrix-parametrization of gauge potentials, analogous
to the form in (\ref{1}).
What we need to do is therefore to obtain a toric version of
(\ref{6}) or (\ref{10}) by use of such a matrix parametrization.
A main claim we like to make in this paper is that, in the
toric case, the level number appeared in (\ref{10})
no longer vanishes since it is now incorporated with
nontrivial zero-mode dynamics. If the toric level number,
say $\tilde{k}$, vanishes, there will be no topological
differences from the planar case. We discuss
contributions of zero modes later in section 4.

%%%%%%%%%%%%%%%%%%%%%%%%%%%%%%%%%%%%%%%%%%%%%%%%%%%%%%%%%%%%%%%%%%%
\section{Matrix parametrization of gauge potentials on torus}

Torus can be described in terms of two real coordinates $\xi_1$, $\xi_2$ with
periodicity of $\xi_i \rightarrow \xi_i + n$ ($i=1,2$) where $n$ is any integer.
Complex coordinates of torus can be
parametrized as $z = \xi_1 + \tau \xi_2$ where $\tau = \re \tau
+ i \im \tau$ is the modular parameter of the torus.
There are two noncontractible cycles on torus, conventionally labeled
as $\al$ and $\bt$ cycles. By use of these a holomorphic one-form of the torus,
$\om = \om(z) dz$, can be defined as
\beq \int_\al \om = 1 ~, ~~~~~~ \int_\bt \om = \tau \label{2-1} \eeq
where the normalization of $\om$ is given by
\beq \int dz d\bz ~ \bom \wedge \om ~=~ i 2 ~ \im \tau \, . \label{2-2} \eeq
Equivalently, this can be written as $\int_\Si \bom \wedge \om = \im \tau$
with $\Si = S^1 \times S^1$.
Note that $\om$ is a zero mode of $\d_\bz$. In construction of
matrix parametrization of gauge potentials on torus, we therefore need to
take  $\om$ and $\bom$ into account.

We shall denote $a$ as a complex physical variable of zero modes for the moment.
We may regard $a$ as an abelian gauge potential corresponding to the
zero modes of torus.
Note that $a$ and $\ba$ satisfy the periodicity
$a \rightarrow a + m + n \tau$ and
$\ba \rightarrow \ba + m + n \bar{\tau}$
($m, n \in {\bf Z}$) where ${\bf Z}$ denotes integer.
$m$ and $n$ correspond to winding numbers of $\alpha$ and $\beta$ cycles,
respectively.

%%%%%%%%%%%%%%%%%%%%%%%%% figure %%%%%%%%%%%%%%%%%%%%%%%%%
\begin{figure} [tbp]
\begin{center}
%\epsfxsize=145mm \epsfysize=45mm
%\epsfbox{fig01.eps}
\includegraphics[width=145mm]{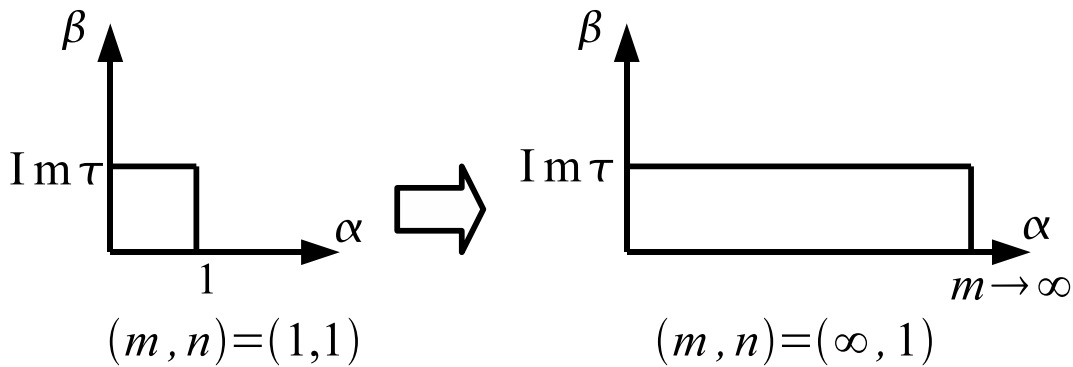}
\caption{Choice of winding numbers for discussion of deconfinement ---
we assume $\re \tau = 0$ and choose $mn$ to be a large integer such that
$m \gg n = 1$.
Note that $\im \tau$ and $m$ can be interpreted as radii for the two circles
of torus $(S^{1}_{\alpha} \times S^{1}_{\beta})$ corresponding to $\alpha$
and $\beta$ cycles, respectively.}
\label{fig1}
\end{center}
\end{figure}
%%%%%%%%%%%%%%%%%%%%%%%%% figure %%%%%%%%%%%%%%%%%%%%%%%%%

Let us consider a change of variables for
the one-forms $\om$ and $\bom$ in terms of $\xi_1$ and $\xi_2$.
Note that $\xi_i$ ($i =1,2$) take real values in $0 \le \xi_i \le 1$ with
 $\xi_i = 0$, $1$ being identical.
Let $\om_1$ and $\om_2$ be
\beqar
\om_1 &=& (d\bz - dz ) / 2i ~=~ -\im \tau d\xi_2 \nonumber \\
\om_2 &=& (\tau d\bz - \bar{\tau} dz) / 2i ~=~  \im \tau d\xi_1
\label{2-3}
\eeqar
where we assume $\re \tau =0$. Since an integral part of $\re \tau$ can be
absorbed into $m$ of the periodicity of $a \rightarrow a + m + n\tau$, this
assumption is equivalent to $\re \tau$ being an integer,
which may not cause obstacles in the following discussion.\footnote{
One might argue the assumption of $\re \tau =0$ is relevant to the imaginary
time formulation as discussed in the introduction but the relevance is not
entirely clear at least for the author.}
With (\ref{2-3}), holonomies of
torus can be rewritten as
\beq \oint_{\al_i} \om_j ~=~  (\im \tau) \, \ep_{ij}
\label{2-4}
\eeq
where $\ep_{ij}$ denotes a Levi-Civita symbol
and $\al_1$, $\al_2$ denote the the alpha and beta cycles
defined in (\ref{2-1}).
Note that we can set
$\om (z) = 1$ in (\ref{2-1}) with identification of the alpha and beta cycles by
loop integrations of the variables $\xi_1$ and $\xi_2$, respectively.
Normalization for $\om_1$ and $\om_2$ is given by
\beq
\int dz d\bz ~ \frac{\om_1}{\im \tau} \wedge \frac{\om_2}{\im \tau} ~=~1 \, .
\label{2-5}
\eeq
We now introduce a new set of variables corresponding to $\om_1$, $\om_2$ by
\beq
a_1 = \ba - a ~~, ~~~ a_2 = \tau \ba - \bar{\tau} a \, .
\label{2-6}
\eeq
Under the transformations of $a \rightarrow a + m + n \tau$ and
$\ba \rightarrow \ba + m + n \bar{\tau}$, $a_1$ and $a_2$ vary as
\beqar
\del a_1 &\rightarrow& (-2i \im \tau) n \, , \nonumber \\
\del a_2 &\rightarrow& (2 i \im \tau) m \, .
\label{2-7}
\eeqar
From (\ref{2-4}) and (\ref{2-7}), we find
\beq
\exp \left(
{\oint_{\al_2} \frac{\pi \om_1}{\im \tau} \frac{\del a_2}{\im \tau}}
\right)
= e^{-i 2 \pi m}  ~, ~~
\exp \left(
{\oint_{\al_1} \frac{\pi \om_2}{\im \tau} \frac{\del a_1}{\im \tau}}
\right)
= e^{-i 2 \pi n} \,.
\label{2-8}
\eeq

For a nonabelian case with an $SU(N)$ gauge group,
physical variables $a$, $\ba$ are given by the following
matrix-valued quantities \cite{GaweKupi1,BosNair2}:
\beq a ~=~ a_j \, t_{j}^{\diag} ~, ~~~
 \ba ~=~ \ba_j \, t_{j}^{\diag}
\label{2-9}
\eeq
where $t_{j}^{\diag}$ are the diagonal generators of $G=SU(N)$ in the fundamental
representation ($j = 1,2,\cdots,N-1)$, corresponding to the Cartan subalgebra
of $G$. $a_j$ are complex variables satisfying
$a_j \rightarrow a_j + m_j + n_j \tau$
with $m_j$ and $n_j$ being integer.
In the expressions of (\ref{2-9}), sums over $j$ should
be understood.

Nonabelian versions of $a_1$ and $a_2$ can also be given by (\ref{2-6})
with $a$ and $\ba$ now defined as (\ref{2-9}).
By use of such $a_1$ and $a_2$, we can express matrix parametrization
of gauge potentials on torus,
which is analogous to the planar case (\ref{1}), as
\beqar
\widetilde{A}_{\xi_1} &=& A_{\xi_1} + M \left(
\frac{\pi \om_2}{\im \tau} \frac{a_1}{\im \tau}
\right) M^{-1} ~,~~~ A_{\xi_1} = -\d_{\xi_1} M \, M^{-1} \nonumber \\
\widetilde{A}_{\xi_2} &=& A_{\xi_2} + M^{\dag -1} \left(
\frac{\pi \om_1}{\im \tau} \frac{a_2}{\im \tau}
\right) M^{\dag}~,~~~ A_{\xi_2} = M^{\dag -1} \d_{\xi_2}  M^{\dag}
\label{2-10}
\eeqar
where $\d_{\xi_i}$ denotes $\frac{\d}{\d \xi_i}$ ($i=1,2$).
Note that there is a set of $\ep_{ij}\om a_j$ combinations appeared in
(\ref{2-8}) such that we have invariance of $\widetilde{A}_{\xi_1}$,
$\widetilde{A}_{\xi_2}$ under the transformations of $(a,\ba)$.
In terms of $a$, $\ba$, the parametrization can be expressed as
\beqar
\widetilde{A}_z &=& - \d_z \widetilde{M} \, \widetilde{M}^{-1}  \nonumber \\
&=& - \d_z M \, M^{-1} + M
\left( \frac{ \pi \om}{\im \tau} \ba \right) M^{-1} \, ,
\label{2-11}\\
\widetilde{A}_\bz  &=& \widetilde{M}^{\dag -1} \d_\bz \widetilde{M}^\dag \nonumber \\
&=& M^{\dag -1} \d_\bz M^\dag + M^{\dag -1}
\left( \frac{ \pi \bom }{\im \tau} a \right) M^\dag
\label{2-12}
\eeqar
with $\widetilde{M}$ and $\widetilde{M}^\dag$ now defined by
\beqar
\widetilde{M} &=& M \exp \left( -
\frac{ \pi }{\im \tau} \int^z \om \ba \right)
\equiv ~M ~ \widetilde{\ga}_z \, , \nonumber\\
\widetilde{M}^\dag &=& \exp \left(
\frac{ \pi }{\im \tau}  \int^\bz \bom a \right) \, M^\dag
\equiv ~ \widetilde{\ga}_\bz ~ M^\dag \, .
\label{2-13}
\eeqar
Equations (\ref{2-11}) and (\ref{2-12}) agree with previously known
matrix parametrization for gauge potentials on torus \cite{GaweKupi1,BosNair2}.

\vspace{.2in}
\noindent
\underline{\emph{A gauge invariant measure and decomposition of $\widetilde{H}$}}

Let us now consider a gauge invariant measure which
incorporates the zero modes of torus. From the calculation of the
planar case shown in (\ref{4}) and from Narashimhan-Seshadri theorem,
we can define the toric measure as
\beq
d \mu (\widetilde{\C}) ~=~
d\mu(\widetilde{H}) ~ e^{2 c_A \S_{WZW}(\widetilde{H})}
\label{2-14}
\eeq
where $\widetilde{H}
= \widetilde{M}^\dag \widetilde{M} = \widetilde{\ga}_\bz
 M^\dag M \widetilde{\ga}_z = \widetilde{\ga}_\bz H \widetilde{\ga}_z$.
The change from $H$ to $\widetilde{H}$
can essentially be taken care of by replacing $\d_\bz$ with
$\d_\bz + \frac{\pi \bom}{\im \tau} a$ and similar for $\d_z$.

Decomposition of $a$, $\ba$ out of $\widetilde{H}$ may be
achieved by imposing
\beq [a , H] ~=~a_j [t_{j}^{\diag} , H]~=~0  ~~~~ (a_j \ne 0)
\label{2-15} \eeq
This is a strong assumption we would like to impose on $H$ later in
the next section. In terms of matrix configuration, this basically leads to
diagonalization of $H$ for arbitrary choices of $a_j$'s ($j=1,2,\cdots, N-1$).
Note that since we are interested in nontrivial zero-mode contributions
we consider non-vanishing $a$ or $\ba$, {\it i.e.}, at least one of $a_j$'s
should be non-zero. For example, we can choose
$a_l = 0$ $(l = 1,2,\cdots, r; \, 1 < r < N-1)$ and
$a_{m} \ne 0$ $(l = r,r+1,\cdots, N-1)$. Then $H$ does not
get fully diagonalized  but has a block-diagonal structure with two blocks,
one of which being an $r$-dimensional block.
Under the assumption of (\ref{2-15}), we can express
the gauge-invariant measure as
\beqar
 \left.d \mu (\widetilde{\C}) \right]_{decom.}&=&
 \left. d\mu( H )  d\mu(a,\ba)  e^{2c_A \S_{WZW}(\widetilde{H})} \right]_{[a,H]=0}
\nonumber \\
&=&  \left. d\mu(\C) d\mu (a, \ba) \right]_{[a,H]=0}
\label{2-16}
\eeqar
where $d\mu(a,\ba) =  \prod_{j=1}^{N-1} d\mu(a_j, \ba_j)$. In the last step,
we neglect $(a, \ba)$-contributions of $\S_{WZW}(\widetilde{H})$, which
may be absorbed into definitions of wave functions.
Since $d\mu (a, \ba)$ is invariant under $a_j \rightarrow a_j + m_j + n_j \tau$,
the expression (\ref{2-16}) shows explicit gauge invariance of the measure
which is incorporated with the zero modes.

%%%%%%%%%%%%%%%%%%%%%%%%%%%%%%%%%%%%%%%%%%%%%%%%%%%%%%%%%%%%%%%%%%%
\section{Zero-mode K\"{a}hler potentials and gauge invariance}

In this section, we construct a wave function corresponding to
the vacuum state of (2+1)-dimensional Yang-Mills theory on
$S^1 \times S^1 \times {\bf R}$.

\vspace{.2in}
\noindent
\underline{\emph{Abelian Case}}

We first focus on abelian zero-mode dynamics and then move to
a nonabelian case later. As in the previous section,
we shall denote $a$ as a complex variable for the moment.
For an abalian case, we can substitute $M^\dag = e^{i \th(z,\bz)}$, where
$\th (z,\bz)$ is a function of $z$, $\bz$, into
(\ref{2-12}) to obtain
$\widetilde{A}_\bz = i \d_\bz \th + \frac{\pi \bom}{\im \tau} a$.
Thus, under a certain gauge, physical variables
of $\widetilde{A}_\bz$ can be given solely by $a$ which
satisfies the periodicity $a \rightarrow a + m + n \tau$.

We now consider geometric quantization of the $U(1)$ Chern-Simons theory,
following the line of \cite{BosNair1,NairBook} in a slightly different manner.
From (\ref{9}), (\ref{2-2}) and (\ref{2-11})-(\ref{2-13}), we can express
the K\"{a}hler form of zero modes as
\beq \Om ~=~ \frac{k_{a\ba}}{2 \pi} da \wedge d\ba \int_{z,\bz}
\left( \frac{ \pi \bom}{\im \tau}  \right)
\wedge \left( \frac{ \pi \om}{\im \tau}  \right)
~=~ i \frac{\pi k_{a\ba}}{\im \tau} da \wedge d \ba
\label{3-1}
\eeq
where the integral is taken over $dz d\bz$ and
$k_{a \ba}$ is the level number associated to the abelian Chern-Simons theory.
The corresponding zero-mode K\"{a}hler potentials can generally be expressed as
\beq
W(a, \ba) =  \frac{\pi k_{a\ba}}{\im \tau} a \ba + g(a) + \bar{g}(\ba)
\label{3-2}
\eeq
where $g(a)$ and $\bar{g}(\ba)$ are purely $a$-dependent and $\ba$-dependent
functions, respectively.

From (\ref{2-2})-(\ref{2-6}) we have a following relation
\beq
da \wedge d \ba \int_{z,\bz} \bom \wedge \om ~=~
da_1 \wedge d a_2 \int_{z,\bz} \frac{\om_2}{\im \tau }\wedge \frac{\om_1}{\im \tau}
\label{3-3}
\eeq
where we use
\beq
\om = \frac{\om_2 - \tau \om_1}{\im \tau} ~,~~
\bom = \frac{\om_2 - {\bar \tau} \om_1}{\im \tau} \,.
\label{3-4}
\eeq
In terms of $a_1$ and $a_2$, the zero-mode K\"{a}hler form
can then be expressed as
\beqar
\Om &=&  \frac{k_{a\ba}}{2 \pi}
\left( \frac{\pi}{\im \tau} \right)^2 da \wedge d \ba
\int_{z,\bz} \bom \wedge \om ~=~ \frac{k_{a\ba}}{2 \pi}
\left( \frac{\pi}{\im \tau} \right)^2 (2i \im \tau) da \wedge d \ba
\nonumber \\
&=& \frac{k_{a\ba}}{2 \pi}
\left( \frac{\pi}{\im \tau} \right)^2 da_1 \wedge d a_2
\int_{z,\bz} \frac{\om_2}{\im \tau} \wedge \frac{\om_1}{\im \tau}
 ~=~ - \frac{k_{a\ba}}{2 \pi}
\left( \frac{\pi}{\im \tau} \right)^2  da_1 \wedge d a_2 \, .
\label{3-5}
\eeqar
A K\"{a}hler potential corresponding to the second line in (\ref{3-5})
may be given by
\beq
K(a,\ba) = \frac{i \pi k_{a \ba}}{2(\im \tau)^2}
(\ba -a)(\tau \ba - \bar{\tau} a )\, .
\label{3-6}
\eeq
$K(a,\ba)$ appears to differ from the general expression $W(a,\ba)$ in
(\ref{3-2}). But this does not cause a problem since
both $W(a, \ba)$ and $K(a,\ba)$ are derived from the same K\"{a}hler form
$\Om$ with different choices of frames, {\it i.e.}, the two K\"{a}hler potentials describe
the same physics of zero modes.
We shall choose $K(a,\ba)$ as our zero-mode K\"{a}hler potential in the following.

The symplectic potential for the zero modes can be expressed as
\beqar
\A &=&  \frac{\pi k_{a\ba}}{4 (\im \tau)^2} \int_{z,\bz}
\left(
\frac{\om_2 a_1}{\im \tau} \wedge \frac{\om_1}{\im \tau} da_2
- \frac{\om_1 a_2}{\im \tau} \wedge \frac{\om_2}{\im \tau} da_1
\right) \nonumber \\
&=& - \frac{\pi k_{a\ba}}{4 (\im \tau)^2}
( a_1 da_2 + a_2 da_1) \, .
\label{3-7}
\eeqar
Note that we take account of the couplings of $a_i$ to $\om_i$ $(i=1,2)$
in $K(a,\ba)$. A naive calculation of $d \A$ with respect to
$a_i$ does not lead to $\Om$ of (\ref{3-5}) but this is not a discrepancy
since $\Om$ is also defined with $a_i$ coupled to $\om_i$. From (\ref{2-7}),
a variation of $\A$ under $a \rightarrow a + m + n \tau$
is given by $\A \rightarrow \A + d \La_{m,n}$ where
\beq
\La_{m,n} ~=~ - i \frac{\pi k_{a\ba}}{2 \im \tau}
( m a_1 - n a_2 )\,.
\label{3-8}
\eeq

A holomorphic wavefunction which satisfies the polarization condition can be
expressed as
\beq
\Psi[\widetilde{A}_{\bz}] \equiv \Psi[a] = e^{-\frac{K(a,\ba)}{2}} f(a) \, .
\label{3-9}
\eeq
We require zero-mode `gauge' invariance on $\Psi[a]$ under
$a \rightarrow a + m + n \tau$ by imposing
\beq e^{i \La_{m,n}} \Psi[a] ~=~ \Psi[a + m + n \tau] \, .
\label{3-10}
\eeq
This leads to the following relation
\beq
f(a) ~=~ e^{- i \pi k_{a\ba} mn} f(a + m + n \tau) \, .
\label{3-11}
\eeq
Since the periodicity property $f(a)=f(a+ m + n\tau)$
is a natural requirement for any functions defined on torus, the relation
(\ref{3-11}) means that $\Psi[a]$ can be `gauge' invariant given that
$f(a)$ satisfies a Dirac-like quantization condition for $k_{a\ba}$, {\it i.e.},
$k_{a\ba} \in 2 {\bf Z}$. This is another indication of level
quantization for the Chern-Simons theory on torus.

For choices of arbitrary winding numbers $(m, n)$, one
may make $n$ be absorbed into $k_{a \ba}$.
This arrows us to identify $2n$ with the level number of abelian
Chern-Simons theory encoding the zero mode dynamics.
We shall later choose $mn$ to be a large integer such that
$m \gg n =1$ as mentioned in Fig.\ref{fig1}.

An inner product of the holomorphic wavefunctions can be expressed as
\beq
\< 1 | 2 \> ~=~
\int d\mu(a,\ba) \, e^{-K(a,\ba)} \, \overline{f_{1}(a)} f_2 (a)
\label{3-12}
\eeq
where $\overline{f_{1}(a)}$ denotes the complex conjugate of
the function $f_{1} (a)$.
Note that, as a requirement for a holomorphic function,
the factor of $\frac{\pi k_{a\ba}}{\im \tau} \ba$ is
realized by an operation of $\frac{\d}{\d a}$ on $f(a)$.
We shall discuss this point later below equation (\ref{5-10}).

\vspace{.2in}
\noindent
\underline{\emph{Nonabelian Case}}

Let us now turn to the main part of the present paper.
From (\ref{8})-(\ref{10}) and Narashimhan-Seshadri theorem,
we can express a wave functional for vacuum states of $(2+1)$-dimensional
Yang-Mills theory on torus as
\beq
\Psi[\widetilde{A}_\bz] ~\equiv~
\Psi[\widetilde{M}^\dag] ~=~
 e^{-\frac{\widetilde{K}}{2}} e^{\tilde{k}
 \S_{WZW}(\widetilde{M}^\dag)} \Upsilon(a)
\label{4-1}
\eeq
where $a$ now has an algebraic structure as in (\ref{2-9}) and $\widetilde{K}$
is a toric version of (\ref{9}), {\it i.e.},
\beq
\widetilde{K} ~=~  -\frac{\tilde{k}}{\pi} \int_\Si
\Tr(\widetilde{A}_\bz \widetilde{A}_z )
\label{4-2}
\eeq
with $\tilde{k}$ being a toric version of the level number $k$ defined in (\ref{9}).
Note that $\Upsilon(a)$ in (\ref{4-1}) is some functions of $a$ which does not depend on
$\widetilde{H} = \widetilde{M}^\dag \widetilde{M}$ as
$\Psi[\widetilde{M}^\dag]$ being a wave functional for the vacuum states.
The inner product can be given by
\beqar
\<1|2\> &=& \int d\mu(\widetilde{\C}) \Psi_{1}^*[\widetilde{M}^\dag]
 \Psi_2 [\widetilde{M}^\dag]  \nonumber \\
&=& \int d\mu(\widetilde{H}) e^{(2 c_A + \tilde{k})  \S_{WZW}(\widetilde{H})}
\overline{\Upsilon_{1}(a)} \Upsilon_{2}(a)
\label{4-3}
\eeqar
where $\overline{\Upsilon_{1}(a)}$ is a complex conjugate of $\Upsilon_{1}(a)$
and $\Psi_{1}^*[\widetilde{M}^\dag]$ is defined by
\beq
\Psi^*[\widetilde{M}^\dag] ~\equiv~
\Psi[\widetilde{M}] ~=~
 e^{-\frac{\widetilde{K}}{2}} e^{\tilde{k}
 \S_{WZW}(\widetilde{M})} \overline{\Upsilon(a)}
\label{4-4}
\eeq
We may naively follow the lines of discussion in the planar case
to assume $\tilde{k} \rightarrow 0$ but this implies ignorance
of nontrivial zero modes; as we will see in a moment,
the level number $\tilde{k}$ can essentially be given by
the zero-mode level number $k_{a\ba}$ in the previous subsection.
To circumvent the problem, we can make a dimensional analysis.
One of the interesting features of Yang-Mills theory in three
dimensions is that a coupling constant $e$ has mass dimension of
$\hf$. Gauge potentials also have $\hf$ mass dimension for
the three-dimensional theory.
Conventionally, this is realized by absorption of $e$
into dimensionless $A_\bz$ in the form of
(\ref{1}).
Similarly, we regard $\im \tau$ in
$\widetilde{A}_\bz$ of (\ref{2-11}) as dimensionless.
Mass dimension of $\im \tau$, however, can be expressed as
$[(\im\tau)^{-1}]=[e^2]=1$.
We can then consider $\frac{\pi}{\im \tau}$ as
a unit of the level number $\tilde{k}$,
where the factor of $\pi$ for $\im \tau^{-1}$ arises from
periodicity conditions (\ref{2-8}) or from the matrix
parametrization (\ref{2-11}), (\ref{2-12}).
In terms of dimensionful parameters, we may explicitly write
the level number $\tilde{k}$
as $\frac{\pi}{\im \tau e^2} \tilde{k}$.
This vanishes as $\im \tau \rightarrow \infty$, which corresponds
to the planar case. For finite $\im \tau$ with nonzero $k_{a\ba}$,
however, the toric level number does not vanish and we have
a critical value of $\frac{1}{ \im \tau}$ at which the factor of
$2 c_A + \tilde{k}$ in (\ref{4-3}) vanishes.
In the KKN Hamiltonian approach, this factor,
or the coefficient of $\S_{WZW}(H)$,
involves in calculations of string tension and mass gap.
We can therefore read off a deconfinement temperature from
the critical value of $\frac{1}{ \im \tau}$ , which we shall
discuss in the next section.

To understand the relation between $\tilde{k}$ and $k_{a\ba}$,
we now consider decomposition of $a$, $\ba$ out of $\widetilde{H}$ with
an imposition of (\ref{2-15}).
The K\"{a}hler potential (\ref{4-2}) can be expressed as
\beqar
 \widetilde{K}_{decom.} &=&
-\frac{\tilde{k}}{\pi} \int_\Si
\Tr(A_\bz A_z )
-\frac{\tilde{k}}{\im \tau} \int_\Si
\Tr(A_\bz M \om \ba M^{-1}
+ M^{\dag -1} \bom a M^{\dag}
A_z ) \nonumber \\
&&~ + \prod_{j}^{N-1} \frac{ \pi \tilde{k}}{2 \im \tau} a_j \ba_j
\label{4-5}
\eeqar
The last term can be written as
$\prod_j \frac{\pi k_{a\ba}}{\im \tau} a_j \ba_j$
with an identification of
\beq
\tilde{k} = 2 k_{a\ba} \, .
\label{4-6}
\eeq
As we have discussed below (\ref{3-11}), we can express
$k_{a\ba} = 2 n$ for any integer $n$. So the identification (\ref{4-6})
means $\tilde{k} = 4 n$.
Note that we may choose a sign of $\tilde{k}$ since
the order of variables does not matter in K\"{a}hler {\it potentials}
of zero modes; this implies a change of sign for (positive)
$k_{a\ba}$ in the definition of zero-mode K\"{a}hler {\it form} in (\ref{3-1}).
As discussed earlier, by a change of frames we may replace
$\frac{\pi k_{a\ba}}{\im \tau} a_j \ba_j$ with
$K(a_j, \ba_j)$ of (\ref{3-6}).
Upon decomposition, the wave functional (\ref{4-1})
can then be written as
\beqar
\Psi[\widetilde{M}^\dag]_{decom.} &=&
e^{-\hf \prod_j K(a_j, \ba_j)} \Upsilon(a)
~ e^{
\frac{\tilde{k}}{2 \pi} \int_\Si \Tr(A_\bz A_z)
}
~ e^{
\tilde{k} \S_{WZW} (M^\dag)
} \nonumber \\
&& ~\times
 \exp \left[
\frac{\tilde{k}}{2 \im \tau} \int_\Si \Tr(A_\bz M \om \ba M^{-1}
+ M^{\dag -1} \bom a M^\dag A_z)
\right]
\nonumber \\
&& ~\times
 \exp \left[
- \frac{\tilde{k}}{\im \tau} \int_\Si \Tr( \bom a \d_z M^{\dag} M^{\dag -1} )
\right]
\label{4-7}
\eeqar
where we use the Polyakov-Wiegmann identity (\ref{11})
for $\S_{WZW}(\widetilde{M^\dag}) = \S_{WZW}(\widetilde{\ga}_\bz M^\dag)$
and
$\d_\bz \widetilde{\ga}_\bz = \frac{\pi \bom}{\im \tau} a
\widetilde{\ga}_\bz$, $\d_z \widetilde{\ga}_\bz = 0$ for
$\widetilde{\ga}_\bz $ defined in (\ref{2-13}).
In analogy with the abelian case, we can now regard $\Upsilon (a)$ as
a general function satisfying invariance under
$a_j \rightarrow a_j + m_j + \tau n_j$ for $a = a_j t_{j}^{\diag}$ and
$m_j , n_j \in {\bf Z}$.
We can have a similar expression for
$\Psi[\widetilde{M}]_{decom.}$ and its product
with $\Psi[\widetilde{M}^\dag]_{decom.}$ can be written as
\beqar
\left.\Psi_{1}^{*} \Psi_2 \right]_{decom.}
&=& e^{- \prod_j K(a_j, \ba_j)}
\overline{\Upsilon_1 (a)} \Upsilon_2 (a)
~ e^{
\tilde{k} \S_{WZW} (H)
} \nonumber \\
&& ~ \times
 \exp \left[
\frac{\tilde{k}}{\im \tau} \int_\Si
\Tr(H^{-1} \d_\bz H \om \ba - \bom a \d_z H H^{-1} )
\right] \, .
\label{4-8}
\eeqar

Earlier we have obtained a gauge invariant wave function for an abelian case
but for nonabelian cases it is not possible to do so. This is because
a corresponding vacuum wave functional
has properties arising from a WZW action.
Note that neither of $\Psi[\widetilde{M}^\dag]$ or $\Psi[\widetilde{M}]$
is gauge invariant in terms of the zero mode variables.
This is related to the fact that there is
no gauge invariant WZW action for $S_{WZW}(\widetilde{M^\dag})=
S_{WZW}(\widetilde{\ga}_\bz M^\dag)$.
Technically speaking, $\S_{WZW}(\widetilde{M^\dag})$ does not
satisfy the so-called anomaly-free condition \cite{Witten1},
which is a sort of chirality condition in terms of transformations
from $M^\dag$ to $\widetilde{M^\dag}$.
On the other hand, $\S_{WZW}(\tilde{H})$  satisfies
the anomaly condition. So we may obtain a gauge invariant
value for the product in (\ref{4-8}).
Indeed, we can introduce the following gauged WZW action
\cite{gWZW1,gWZW2,Witten1}:
\beqar
I(H, a) &=& \S_{WZW} (H) + \frac{1}{\pi} \int_\Si \Tr \left(
H^{-1} \d_\bz H \frac{\pi \om}{\im \tau} \ba ~-~
\frac{\pi \bom}{\im \tau} a \d_z H H^{-1} \right. \nonumber \\
&& \hspace{4.0cm}
\left. ~+~ H^{-1} \frac{\pi \bom}{\im \tau} a H \frac{\pi \om}{\im \tau} \ba
~-~ \frac{\pi \bom}{\im \tau} a \frac{\pi \om}{\im \tau} \ba \right) \, .
\label{4-9}
\eeqar
The inner product is then given by
\beq
\< 1 | 2 \>_{decom.} ~=~ \left. \int d\mu(H) d\mu(a, \ba)
e^{(2c_A + \tilde{k}) I(H,a)}
~
\overline{\Psi_1 (a)} \Psi_2 (a) \right]_{[a, H]=0}
\label{4-10}
\eeq
where $\Psi(a)$ is now defined as
\beq
\Psi(a) ~=~ \exp \left(
- \prod_{j}^{N-1} \frac{K(a_j, \ba_j)}{2} \right)
\Upsilon (a) \, .
\label{4-11}
\eeq
$\overline{\Psi(a)}$ is a complex conjugate of $\Psi(a)$.
$\Upsilon (a)$ is a general function on torus with a Cartan subalgebra
structure for $a$.
Equation (\ref{4-10}) shows manifest gauge invariance
of the inner product in terms of $H$ and $(a, \ba)$,
including the measure.

What we have done here is to obtain a lower class of the
vacuum wave functional (\ref{4-1}) and a corresponding
inner product by imposition of the decomposition assumption
(\ref{2-15}). The decomposition condition
leads to expressions in terms of $H$ and $(a,\ba)$ by `gauging away'
$(a,\ba)$-dependence in $\widetilde{H} = \widetilde{\ga}_\bz
H \widetilde{\ga}_z$ so that we can extract the gauge invariant matrix $H$.
In (\ref{4-8}) we find a coupling between the current of
$S_{WZW}(H)$, $\d_z H H^{-1}$, and the zero mode variable
$\bom a$. This coupling implies an identification of
the current as a non-perturbative gluon field as discussed
in the Hamiltonian approach. This is an interesting point,
however, the upshot of the decomposition analysis here lies in
gauge invariance of the inner product which leads to the relation of
level number $\tilde{k}$ with $k_{a\ba}$ as in (\ref{4-6}) and
properties of $\Upsilon (a)$ as a general non-abelian function on torus.
Since $a = a_j t_{j}^{\diag}$
is diagonal, the decomposition condition (\ref{2-15}) does not
affect on the properties of $\Upsilon (a)$; hence, that in (\ref{4-1})
remains the same as a Cartan subalgebraic function on torus throughout
the present subsection.

%% additional explanations
The expression (\ref{4-10}) suggests that the effects of zero modes
on torus can be interpreted as changes in level numbers of the WZW action.
In this sense, $(2+1)$-dimensional Yang-Mills theory on
$S^1 \times S^1 \times {\bf R}$ may be regarded as Yang-Mills-Chern-Simons theory.
This is, however, not in contradiction to our consideration of pure Yang-Mills
theory because of the following. As mentioned in the introduction, physical states of
$(2+1)$-dimensional Yang-Mills theory can be described by holomorphic wave functionals
of Chern-Simons theory.
The apparent changes of level numbers in
the WZW action arise from the zero-mode contributions
to the holomorphic wave functionals of Chern-Simons theory in the toric case.
Thus we are considering pure Yang-Mills theory on $S^1 \times S^1 \times {\bf R}$,
with the zero-mode contributions represented by $\tilde{k}$ in the exponent of
(\ref{4-10}) along with an additional abelian measure $d \mu (a , \bar{a} )$.

\vspace{.2in}
\noindent
\underline{\emph{Vacuum states and theta functions}}

So far we have chosen a K\"{a}hler potential of either abelian or
nonabelian theory such that a holomorphic function, $f(a)$ or $\Upsilon (a)$,
has a obvious periodic relation characterized by (\ref{3-11}).
We can however use different K\"{a}hler potentials to start with, as long
as the potentials lead to the same K\"{a}hler form.
For example, it is known a certain choice of K\"{a}hler potential
gives rise to theta functions for holomorphic functions for an
abelian case \cite{BosNair1,NairBook}.
We shall briefly review this fact in the
rest of this section for better understanding of the above mentioned results
in the framework of geometric quantization.

A K\"{a}hler potential we choose is
\beq
W_\tau (a, \ba) = \frac{i \pi k_{a\ba}}{2 (\im \tau)^2}(\ba - a)^2 \tau
\label{5-1}
\eeq
which, along with an implicit assumption of $\re \tau =0$, can be
expressed in the form of (\ref{3-2}) and leads to the K\"{a}hler form (\ref{3-1}).
A polarization condition for a holomorphic function $\Psi_\tau (a)$ is
given by $\D_\ba \Psi_\tau (a) = ( \d_\ba - i \A_\ba) \Psi_\tau (a)= 0$
where $\A_\ba = \frac{i}{2}\d_\ba W_\tau (a,\ba)$. The holomorphic function
is then given by
\beq
\Psi_\tau (a) = \exp \left[{\frac{- i \pi k_{a \ba}}{4 (\im \tau)^2}(\ba -a )^2 \tau}
\right] f_\tau (a)
 \, .
\label{5-2}
\eeq
This is a general expression for a holomorphic functions upon a choice
of the K\"{a}hler potential as we have seen earlier.
Properties of a holomorphic function $f_\tau (a)$ can similarly be obtained
by imposing gauge invariance on $\Psi_\tau (a)$. With a choice of the
corresponding symplectic one-form of the form \cite{BosNair1,BosNair2}:
\beq
\A_\tau ~=~ -\frac{\pi k_{a\ba}}{2(\im \tau)^2} (\ba - a)
(\tau d\ba - \bar{\tau} da)
\, ,
\label{5-3}
\eeq
we have $\del \A_\tau = d \La_{\tau \, m,n}$ for transformations of
$a \rightarrow a + m + n \tau$ with
\beq
\La_{\tau \, m,n} ~=~ \frac{i \pi k_{a\ba}}{\im \tau} n (\tau \ba - {\bar \tau}a)
\, .
\label{5-4}
\eeq
Gauge invariance on $\Psi_\tau (a)$ is then given by
$e^{i \La_{\tau \, m,n}} \Psi_\tau (a) = \Psi_\tau (a + m + n \tau)$. This leads
to the following relation
\beq
f_\tau (a) = e^{
i2\pi k_{a\ba} \left( \frac{n^2}{2}\tau + a n \right)
} f_\tau  (a + m + n \tau)
\, .
\label{5-5}
\eeq
This shows that $f_\tau (a)$ is a Jacobi $\theta$-function
defined by
\beq
\theta (a, \tau) =
\Theta \left[
    \begin{array}{c}
        0 \\
        0 \\
    \end{array}
\right] (a, \tau)
\label{5-6}
\eeq
where
\beq
\Theta \left[
    \begin{array}{c}
        a \\
        b \\
    \end{array}
\right] (z, \tau) = \sum_{n \in {\bf Z}}
e^{i \pi k_{a\ba} \tau (n+a)^2 + 2 i\pi k_{a\ba}(n+a)(z+b)} \,.
\label{5-7}
\eeq

An operation of $\frac{\d}{\d a}$ on
$f_\tau (a)$ corresponds
$\frac{\pi k_{a\ba}}{\im \tau} (\ba - a)$.
acting on $f_\tau (a)$.
This is in consistent with the K\"{a}hler form written by
$\Om = -i \frac{\pi k_{a\ba}}{\im \tau} d(\ba - a) \wedge da$.
In terms of $f_\tau (a)$, the inner product for the holomorphic functions
is expressed by
\beq
\< 1 | 2 \> = \int d\mu(a, \ba) ~e^{-W_\tau (a,\ba)}~
\overline{f_{\tau 1} (a)} f_{\tau 2} (a) \, .
\label{5-8}
\eeq
Expanding the K\"{a}hler potential, we can rewrite this as
\beq
\< 1 | 2 \> = \int d\mu(a, \ba)~ e^{-\frac{\pi k_{a\ba}}{\im \tau} a \ba }
~ \overline{g_{\tau 1} (a)} g_{\tau 2} (a)
\label{5-9}
\eeq
where we introduce
\beq
g_\tau (a) ~=~ \exp \left[ \frac{\pi k_{a\ba} a^2}{2 \im \tau} \right]
f_\tau (a) \, .
\label{5-10}
\eeq
We now find the operation of $\frac{\d}{\d a}$ on $g_\tau (a)$ is
realized by $\frac{\pi k_{a\ba}}{\im \tau} \ba$.
We further find that
\beq
g_\tau (a) = g_\tau (a + m + n \tau)~~~~~~ (m=0, \, \re \tau =0)
\label{5-11}
\eeq
regardless a choice of $k_{a\ba}$. This periodic relation is a
subsector of the relation for more general holomorphic function $f(a)$ in
(\ref{3-11}) since relation (\ref{5-11}) is realized when $m=0$ and
$\re \tau = 0$ are satisfied. The choice of $f(a) = g_\tau (a)$
is then a concrete realization of the relation
\beq
\frac{\d}{\d a} ~f(a) ~=~ \frac{\pi k_{a\ba}}{\im \tau} \ba ~ f(a) \, .
\label{5-12}
\eeq
What is essential in construction of wave functions of zero modes and
a corresponding inner product is the K\"{a}hler form to start with.
Different K\"{a}hler potentials may
lead to different expressions, {\it e.g.}, (\ref{5-8}) and (\ref{5-9}),
yet physical consequences should be unaltered. With such a principle,
we may require the relation (\ref{5-12}) for
the previously discussed holomorphic function $f(a)$ in (\ref{3-9}).

Extension to a nonabelian case is straightforward with a knowledge of
$SU(N)$ algebra. The theta function is to be replaced by a higher
dimensional theta function, more precisely, the Weyl-Kac character
for $SU(N)$ algebra with level number $k_{a\ba}$ \cite{CFT}:
\beq
ch_{\hat{\la}}(a, \tau) ~=~ \Tr_{\hat{\la}}~
e^{ \pi i k_{a\ba} \tau h^2 - 2 \pi i k_{a\ba}
(a_1 h_1 + a_2 h_2 + \cdots + a_{N-1} h_{N-1} ) }
\label{5-13}
\eeq
where $h = h_1 + h_2 + \cdots + h_{N-1}$
and the trace means
\beq
\Tr_{\hat{\la}} =
\sum_{h \in {\bf Z} + \frac{\la}{k_{a\ba}} } \, .
\label{5-14}
\eeq
Note that for vacuum wave functions, we should take
the ground state for $\hat{\la}$, {\it i.e.}, $\la = 0$.
A nonabelian version of a vacuum wave functional
$\Psi_0 [\widetilde{A}_\bz]$ can be constructed by use of (\ref{5-13}).
Such a wave functional has been studied before
and is explicitly given in \cite{BosNair2}.

%%%%%%%%%%%%%%%%%%%%%%%%%%%%%%%%%%%%%%%%%%%%%%%%%%%%%%%%%%%%%%%%%%%
\section{Deconfining limit}

In this section, we return to a physical part of the present paper.
We consider contributions of the zero modes to the
planar case by taking the winding numbers in the limit
of $(m,n)=(\infty, 1)$.
Since the zero-mode contribution is described by
Chern-Simons theory on torus with level number $\tilde{k}$,
the effect of zero modes can be evaluated with a replacement
of $2c_A$ with $(2 c_A + \tilde{k})$
as seen in equation (\ref{4-3}) (see also \cite{YMCS}, for
rigorous discussion).
From earlier discussion below equation (\ref{4-4}),
we can express $\tilde{k}$ as
$- \frac{2 \pi k_{a\ba}}{\im \tau e^2}$ with $k_{a\ba}
= 2 n$. The critical temperature is therefore
given by
$\left(\frac{1}{\im \tau}\right)_{c} = \frac{e^2 N}{2 \pi}$
where we use $c_A = N$ and $n=1$.
The deconfinement temperature is then expressed as
\beq
T_c = \frac{e^2 N}{2 \pi}
\label{6-1}
\eeq
which is the same as a mass for non-perturbative gluons predicted in
the KKN Hamiltonian approach. Thus $T_c$ in (\ref{6-1}) is
a natural result and it is what we seek for in the present study.
In what follows,
we shall briefly review the calculation of string tension
in the Hamiltonian approach for completion of our discussion.

The vacuum expectation value of the Wilson
loop operator $\< W(C) \>_0 = \< \Psi_0 | W(C) | \Psi_0 \>$
can be calculated as
\beq
\< W(C) \>_0 ~=~ \int d\mu(\widetilde{H})~
e^{(2c_A + \tilde{k}) \S_{WZW} (\widetilde{H})} e^{- S(\widetilde{H})}
~\overline{\Upsilon (a)}\Upsilon (a)~ W(C)
\label{6-2}
\eeq
where the Wilson loop operator is given by
$W(C) = \Tr {\rm P} \exp \left(- \oint \widetilde{A} \right) =
\Tr {\rm P} \exp \left(\frac{\pi}{c_A} \oint \widetilde{J} \right)$
with
$\widetilde{J} = \frac{c_A}{\pi} \d_z \widetilde{H}
\widetilde{H}^{-1}$.
The function $S(\widetilde{H})$ denotes a contribution
from the potential energy of the Yang-Mills theory.
For modes of low momenta, or for a (continuum) strong coupling limit,
this function can be evaluated. Using an analog of two-dimensional Yang-Mills
theory and setting $\Upsilon (a) =1$, we can
evaluate the vacuum expectation as
\beq
\< W(C) \>_0 ~\approx~ \exp \left[ - \tilde{\si} \A_C \right]
\label{6-3}
\eeq
where $\A_C$ is the area of the loop $C$ and $\tilde{\si}$ is the
string tension on torus given by
\beq
\tilde{\si} ~=~ \frac{e^4 }{4 \pi} \left(c_A + \hf \tilde{k}
\right) c_F \, .
\label{6-4}
\eeq
Here $c_F = (N-1)(N+1)/2N$
is the quadratic Casimir for $SU(N)$ in the fundamental representation.
Substituting $\tilde{k} = - \frac{2 \pi k_{a\ba}}{\im \tau e^2}$
($k_{a\ba} = 2 n$, $n = 1$), we find
vanishing of the string tension at $T_c$.

Temperatures corresponding to $n > 1$ are irrelevant in our setting.
However, we can in fact choose a finite $n \ne 1$ as long
as $m \gg n \ge 0$ is satisfied and in this case $\im \tau$
is scaled to $n \im \tau$ such that ${\tilde k}$ remains the
same for any $n$. Notice that for the case of $n=0$,
which may be possible as we consider $n$ as the winding number
of the beta cycle of torus, we have vanishing of $n \im \tau$
but yet $\tilde{k}$ remains as $\tilde{k} = - \frac{4 \pi}{\im \tau e^2}$.
In terms of the picture in Fig.\ref{fig1}, the choice of $n=0$ means
that the torus of our interest is dimensionally reduced
to a circle (times a point).
Thus, in our setting the choice of $n=0$ may be ruled out
for a dimensional reasoning.

In the planer case, the string tension $\si$ is given by
\beq
\si ~=~ e^4 \left( \frac{N^2 - 1 }{8 \pi}\right) \, .
\label{6-6}
\eeq
Comparisons with numerical data can be made for dimensionless
parameter $T_c / \sqrt{\si}$. Our prediction for this value is
\beqar
\frac{T_c}{\sqrt{\si}} &=&
 \sqrt{\frac{2}{\pi}}\sqrt{\frac{N^2}{N^2 -1}} \nonumber \\
 &=& 0.798 \sqrt{\frac{N^2}{N^2 -1}} \, .
\label{6-7}
\eeqar
Lattice simulations show 0.865, 0.903 and 0.86(7) for this value at
$N \rightarrow \infty$. These values are taken from
references \cite{Teper2}, \cite{Teper3} and \cite{Narayanan2}, respectively.
Corresponding error percentages are 8.40\%, 13.2\% and 8.65\%.
Among the lattice data, the one given by \cite{Teper3}
actually provides the most updated and reliable result.
Obviously, we need to make further investigation to figure out
the relatively large deviation between the lattice data
and the value (\ref{6-7}).

%%%%%%%%%%%%%%%%%%%%%%%%%%%%%%%%%%%%%%%%%%%%%%%%%%%%%%%%%%%%%%%%%%%
\section{Concluding remarks}

In the present paper, we consider $(2+1)$-dimensional Yang-Mills theory
on $S^1 \times S^1 \times {\bf R}$ in the framework of the so-called
Karabali-Kim-Nair (KKN) Hamiltonian approach.
A physical motivation to consider the toric theory is clear
since we may regard it as the planar theory at a finite temperature
in the limit of a large radius for one of the $S^1$'s of torus ($S^1
\times S^1$), and, hence, we can discuss deconfinement transition in terms of the other
radius. In order to execute a calculation of a deconfinement temperature,
however, we need to understand some mathematical aspects of the KKN
Hamiltonian approach. In section 2, we review few features of
the Hamiltonian approach which are pertinent to our discussion.
Detailed analysis on dynamics or geometry of zero modes of torus
is given in section 3 for both abelian and nonabelian cases.
For a nonabelian case, we construct vacuum-state wave functionals
for $(2+1)$-dimensional Yang-Mills theory on $S^1 \times S^1 \times {\bf R}$
by use of Narashimhan-Seshadri theorem.
We further consider a subsector of the vacuum wave functionals by imposing
a certain condition (\ref{2-15}) to discuss gauge invariance of
an inner product of the vacuum wave functionals.
Along the way, we also find zero-mode contributions to the planar Yang-Mills theory.
In section 4, we compute a string tension of pure Yang-Mills
theory on $S^1 \times S^1 \times {\bf R}$ in the Hamiltonian framework,
namely, in the so-called continuous strong coupling limit,
and find a deconfinement temperature (\ref{6-1}). This value
agrees with numerical data from lattice simulation
quite roughly in 10\%. We shall leave the explanation of this rather
large error for future studies.

Now we would like to comment on subtle points in the arguments
which we have used in the present paper.
We argue that we take a cylindrical limit
of the torus, {\it i.e.}, $S^1 \times S^1 \times {\bf R}
\rightarrow S^1 \times {\bf R}^2$,
in the end of calculations.
In our framework, one of the $S^1$ directions of the torus
corresponds to the time coordinate. The Lorentz invariance
however implies that we can interchange this temporal direction
with the spatial direction ${\bf R}$.
Thus our results suggest that a change of topologies (from
plane to cylinder) leads to an apparent change of level numbers in a WZW model
which is relevant to the gauge invariant measure in $(2+1)$-dimensional
Yang-Mills theory.
This is a nontrivial result and probably contains some subtleties
because that a topology change causes a change of level numbers
is simply counter-intuitive.
In this paper, we present one of the justifications of this issue
by use of gauge invariance. We have made the following argument.

In the KKN Hamiltonian approach,
the Gauss law constraint (or the integrability) of the gauge potentials
should be satisfied regardless what Riemann surfaces we use in
construction of the WZW action which is relevant to the
gauge invariant measure of $(2+1)$-dimensional Yang-Mills theory.
Thus, the wave functionals in the toric theory can
be written as (\ref{4-1}).
(This is why the use of Narashimhan-Seshadri
theorem has been emphasized in the present paper.)
The symplectic structure of the zero modes is essentially
given by (\ref{3-5}). This is not exactly the symplectic structure
of Chern-Simos theory. However, with an algebraic extension (\ref{2-9}),
we can relate the level number $k_{a \bar{a}}$ of (\ref{3-5}) to
the level number $\tilde{k}$ of (\ref{4-1}) so that
we can encode the zero-mode contributions in the level number $\tilde{k}$.
A detailed explanation of this relation is given
by the argument of gauge invariance in section 4.
The argument is limited to a particular case, where we can
explicitly write down the gauge invariant measure in terms of zero modes.
Although this will be sufficient to show the relation for our purposes,
it is desirable to confirm this relation in more general cases.
We shall leave this task for future studies.

Lastly, we would like to emphasize that the validity of
our analyses and results is limited in the framework of
the KKN Hamiltonian approach.
It is within this framework that we can properly use
the abelian Chern-Simons symplectic form to discuss
zero-mode contributions to $(2+1)$-dimensional Yang-Mills theory.
We have not proven the use of the Chern-Simons symplectic form in general.
Furthermore there may be some subtleties to justify
this analysis in the more standard approaches to Yang-Mills theory.
Although this might be the case, what
is significant in the present paper from a physical perspective is
that the use of the Chern-Simons symplectic form does seem
to lead a reasonable estimate for the deconfinement temperature
and that this fact itself suggests the usefulness of the
KKN Hamiltonian approach in the future investigations of
$(2+1)$-dimensional Yang-Mills theory.

%%%%%%%%%%%%%%%%%%%%%%%%%%%%%%%%%%%%
\vskip .3in
\noindent
{\bf Acknowledgments} \vskip .06in\noindent
The author would like to thank Professor V.P. Nair for
introduction to the present work and helpful discussions
in the spring of 2006. Later correspondence with
Professors D. Karabali and V.P. Nair was of significant help in improving a manuscript.
The author thanks the Yukawa Institute for Theoretical Physics at Kyoto University.
Discussions during the YITP workshop YITP-W-07-05 on
``String Theory and Quantum Field Theory'' were useful for the present work.
The author also thanks Professor Holland for updated information
on lattice results. Lastly the author is grateful to the referee of this paper for many
critical and stimulating comments.

%%%%%%%%%%%%%%%%%%%%%%%%%%%%%%%%%%%%%%%%%%%%%%%%%%%%%%%%%%%%%%%%

\end{document}